\documentstyle[prl,aps,multicol]{revtex}
\renewcommand{\narrowtext}{\begin{multicols}{2}
\global\columnwidth20.5pc}
\renewcommand{\widetext}{\end{multicols} \global\columnwidth42.5pc}

\begin{document}
\draft
\title{On the relationship between the noise-induced persistent
current and dephasing rate.}
\author{V.E.Kravtsov}
\address{The Abdus Salam International Centre for Theoretical Physics,
P.O.B.
586, 34100 Trieste, Italy, \\ Landau Institute for Theoretical
Physics, 2 Kosygina st., 117940 Moscow, Russia, and \\ Centre for Advanced
Study, Drammensveien 78, N-0271 Oslo, Norway.}
\author{B.L.Altshuler}
\address{NEC Research Institute, 4 Independence Way,
Princeton, NJ 08540, USA and \\Princeton University, Princeton, NJ 08545,
USA}
\maketitle
\begin{abstract}
 
  AC noise in disordered conductors causes both dephasing of the
  electron wave functions and a DC current around a mesoscopic
  ring. We demonstrate that the dephasing rate
  $\tau_{\varphi}^{-1}$ in long wires and the DC current, induced
  by the same noise and averaged over an ensemble of small rings
  are connected.  The ensemble-averaged $h/2e$ flux harmonic
  $\langle I\rangle$ of the current and the dephasing rate caused
  by the same uniform in space high frequency AC field are
  related in a remarkably simple way: $\langle I
  \rangle\,\tau_{\varphi}=C_{\beta}\,e$. Here $e$ is an electron
  charge, and the constant $C_{\beta}$ depends on the Dyson
  symmetry class.  For a pure potential disorder the current
  $\langle I\rangle$ is diamagnetic $C_{\beta}=-(4/\pi)$ and in
  the presence of strong spin-orbit scattering it is paramagnetic
  $C_{\beta}=(2/\pi)$.  The relationship seems to agree
  reasonably with experiments.  This suggests that the two
  puzzles: anomalously large persistent current [L.P.Levy et al.,
  Phys.Rev.Lett., {\bf 64}, 2074 (1990)] and the low-temperature
  saturation of the dephasing [P.Mohanty et al., Phys.Rev.Lett.,
  {\bf 78}, 3366 (1997)] may have a common solution.
\end{abstract}

\pacs{PACS number(s): 72.15.Rn, 72.70.+m, 72.20.Ht, 73.23.-b}
\narrowtext

Since the discovery of universal conductance fluctuations
\cite{LeeSt,Alt} physics of mesoscopic systems has made
tremendous progress.  However, few important experimental
observations still remain unexplained.  One of the long-standing
challenges is the anomalously large value of persistent current
\cite{BIL}. Levy et al.  \cite{Levy} studied the magnetic
response of an ensemble of $10^{7}$ mesoscopic copper rings as a
function of applied weak magnetic field. The average current per
ring $\langle I_{0}(\phi)\rangle = I_{0}\sin(4\pi e\phi/hc)$
found from such measurements is a $hc/2e$-periodic function of
the magnetic flux $\phi$ threading each ring. The amplitude
$I_{0}$ has been found to be of the order of $I_{0}^{exp}\sim e/
\tau_{D}$, where $\tau_{D}=L^2/D$ is the time of diffusion around the ring
of the circumference $L$.  Other measurements
\cite{WW,MohW,Bouch} of persistent current
brought up similar results.

On the other hand, assuming that {\it the system is in the
  equilibrium} and that electrons do not interact with each
other, one gets \cite{AGef} the amplitude $I_{0}^{theor}\sim
(e/h)\,\Delta=g^{-1}\,I_{0}^{exp}$, where $\Delta$ is the
electron mean level spacing, and $g= \tau_{D}\Delta$ is the
dimensionless conductance.  The sign of the average persistent
current predicted by the existing theories of non-interacting
electrons, as well as the sign of the contribution of the
electron-electron interaction in non-superconducting systems, is
{\it paramegnetic}, i.e., $\partial \langle I_{0}(\phi) \rangle/
\partial \phi >0$ at small $\phi$.

In the experiments Ref.\cite{Levy,WW,MohW} the dimensionless
conductance was large $g\sim 10^{2}$, i.e., the observed
persistent current exceeded the theoretical estimation by two
orders of magnitude. None of numerous attempts (see e.g.
\cite{AE} and for discussion \cite{Altsh}) succeeded in
explaining the magnitude of the persistent current by
electron-electron interaction.  Therefore, the amplitude of the
persistent current is in striking disagreement with existing
theories.

Another major puzzle in mesoscopic physics recently attracted
much attention.  Mohanty et al.  \cite{Mohan} experimentally
proved that the saturation of the dephasing rate at low
temperatures T cannot be explained by conventional arguments such
as magnetic impurities or heating.  On the other hand, the
dephasing due to interactions between electrons (or between
electrons and phonons) in an equilibrium system is theoretically
predicted to disappear at T=0 \cite{AAK,Alein}.  Several attempts
have been made to resolve the puzzle resulting from the noise
caused by the two-level systems \cite{Imry}, 2-channel Kondo
effect \cite{Zaw}, or external radiation \cite{AltGersh}.  In all
these explanations electrons interact with an `environment' that
displays a {\it real} time evolution, e.g., real transitions in
two-level systems or a time-dependent electric field.  From this
point of view, the system of free electrons subject to an
external AC field captures all the essential features of
dephasing.  For instance using the proper correlator of the equilibrium
{\it intrinsic} AC electric noise one can evaluate \cite{AAK} the
dephasing rate due to $e-e$ interaction, which is in agreement
with the experiment at modestly low temperatures.  \cite{Alein}.

It is known \cite{FKh} that the AC electric field may also cause
a random DC current in mesoscopic systems. However, in contrast
to classical physics where the rectification exists only in media
without inversion center, the Aharonov-Bohm effect makes the {\it
  disorder-averaged} rectified current also possible \cite{KY}.
This rectified DC current leads to the DC magnetic response
similar to the one which results from the equilibrium persistent
current.

We show below that such a noise-induced DC current averaged over
an ensemble of {\it small} rings of the circumference $L\ll
L_{\varphi}$ and the dephasing rate induced by the same noise in
{\it long} wires of the length ${\cal L}\gg
L_{\varphi}=\sqrt{D\tau_{\varphi}}$ are related in a remarkably
simple way:
\begin{equation}
\label{WF}
I_{E}\,\tau_{\varphi}=
C_{\beta}\,\,e,\;\;\;\;C_{\beta}=\left\{\matrix{
-4/\pi, & \beta=1\cr +2/\pi, &\beta=4\cr}
\right.
\end{equation}
where $e$ is an absolute value of the charge carriers and
$C_{\beta}$ is a constant that depends on the Dyson symmetry
class: $\beta=1$ for the pure potential disorder and $\beta=4$ in
the presence of a spin-orbit scattering with the characteristic
length $L_{so}\ll L$.  Thus, the important differences between
the equilibrium persistent current and the rectified DC current
are (1) the magnitude and (2) the sign if $L_{so}\gg L$.  In this
orthogonal case the ensemble-averaged DC current $\langle
I_{E}(\phi)\rangle=I_{E}\,\sin(4\pi e\phi/hc)$ is {\it
  diamagnetic}, i.e. $\partial \langle
I_{E}(\phi)\rangle/\partial \phi <0$ at small $\phi$.

This is the central result of the Letter.  
The relationship Eq.(\ref{WF}) holds regardless of the nature of
the noise, since no single parameter of the system and
`environment' enters Eq.(\ref{WF}). The noise could be external
or intrinsic and not even necessarily electric (e.g., the phonon
wind).  The only important condition is that the noise must be
{\it non-equilibrium}.  This suggests (see also Ref.\cite{Mo})
that the two puzzles: an anomalously large persistent current and
an anomalously large temperature-independent dephasing rate maybe
closely related.

Actually, the relationship Eq.(\ref{WF}) can be understood by the
dimension analysis.  Consider as an example of a noise a
monochromatic AC electric field with a frequency $\omega$ and an
amplitude $E_{\omega}$. Given the diffusion constant $D$ one can
construct a dimensionless combination:
\begin{equation}
\label{al}
\alpha=\frac{D}{\omega^3}\,\left(\frac{e}{h}\,E_{\omega} \right)^2
=\left(\frac{L_{\omega}}{L_{E}} \right)^2,
\end{equation}
where $L_{\omega}=\sqrt{D/\omega}$, and the characteristic
length $L_{E}$ is determined by the equation $e
E_{\omega}L_{E}=h\omega$.  One can estimate the dephasing rate in
{\it long} wires at $T=0$ as:
\begin{equation}
\label{tau}
\frac{1}{\tau_{\varphi}}=\omega\,f_{\varphi}(\alpha).
\end{equation}
Evaluation of $f_{\varphi}(\alpha)$ goes beyond the dimension
analysis.

As to the nonlinear DC current in {\it mesoscopic} rings, its
amplitude depends on {\it two} parameters:
\begin{equation}
\label{fI}
I_{E}=e\,\omega\, f_{I}(\alpha,\gamma),
\end{equation}
- the parameter $\alpha$ Eq.(\ref{al}) and the  `mesoscopic' parameter
\begin{equation}
\label{gamma}\gamma=\omega \tau_{D}=\left(\frac{L}{L_{\omega}}
\right)^{2}.
\end{equation}

In the weak-field limit $\alpha\rightarrow 0$ both the DC current
and the dephasing rate are quadratic in $E_{\omega}$, i.e., linear
in $\alpha$:
\begin{equation}
\label{weak}
I_{E}= e\, \omega\, \alpha\, f_{I}(\gamma),\;\;\;
\tau_{\varphi}^{-1}=\omega
f_{\varphi}'(0)\,\alpha
\end{equation}
where $f_{I}(\gamma)$ is yet an unknown function of $\gamma$.
Provided that this function has a non-zero limit $f_{I}(\infty)$
at $\gamma\gg 1$, we immediately arrive at:
\begin{equation}
\label{WF1}
I_{E}\,\tau_{\varphi}=e\,
C_{\beta},\;\;\;\;C_{\beta}=f_{I}(\infty)/f_{\varphi}'(0),
\end{equation}
where $C_{\beta}$ is a constant of order 1.

This is essentially Eq.(\ref{WF}). The above analysis suggests
that Eq.(\ref{WF}) is valid when $\alpha$ is small, and $\gamma$
is large. According to Eqs.(\ref{al}-\ref{gamma})
$L_{\varphi}\sim L_{\omega}\,\alpha^{-1/2}\sim
L\,(\alpha\gamma)^{-1/2}$.  Thus, the mesoscopic condition $L\ll
L_{\varphi}$ is equivalent to $\alpha\gamma\ll 1$, and the
above consideration is valid when
\begin{equation}
\label{condit}
1\ll \gamma\ll \alpha^{-1},\;\;\;\; L_{\omega}\ll L \ll L_{\varphi}.
\end{equation}

\noindent One can also write Eq. (8) as
$\omega\gg$max$\{\tau_{D}^{-1},eE_{\omega}L/\hbar\}$.

Another condition concerns the space correlation of the field
$E_{\omega}$.  We neglected space dependence of the field.  This
can be done \cite{KY} if at the length
scale of $L_{\omega}$ the field is strongly correlated.

An assumption that $f_{I}(\gamma)$ has a finite limit at
$\beta\rightarrow\infty$ is anything but trivial.  It 
implies that in the quadratic in $E_{\omega}$ approximation the
DC current flows coherently even at $L_{\omega}\ll L$. It was
first mentioned in Ref.\cite{FKh} and further discussed in
\cite{KY,KY1} that the nonlinear DC current is not destroyed at
$L_{\omega}\ll L$.  Note that this conclusion applies only to the
DC current and is {\it not correct} for the ensemble-averaged second
harmonic
current \cite{KY}.

It is intuitively clear that for DC current $I_{E}$  to flow,
the environment and the electrons
{\it should be out of the thermal equilibrium}.
Indeed, $I_{E}$ vanishes identically for the equilibrium electric noise
\cite{KY1}.
On the other hand, even the equilibrium electric noise causes
dephasing \cite{AAK}.

Eq.(\ref{WF}) involves $T \rightarrow 0$ limits of
$1/\tau_{\varphi}$ and the DC current, which are their maximal
values for a given sample. At finite temperatures, the dephasing
rate exceeds $I_{E}/C_{\beta}\,e$ even at $L \ll L_{\varphi}$,
due to the T-dependent contribution from the equilibrium part of
the noise to $1/\tau_{\varphi}$.

Of course, the arguments presented above cannot substitute an
analytic derivation which we proceed with.  Consider a quasi-1D
system of non-interacting electrons with an external AC field
$E(t)=-\frac{1}{c}\,\frac{\partial A_{t}}{\partial t}$ where
$A_{t}$ is a time-dependent tangential vector-potential with the
zero mean value $\overline{A_{t}}=0$. Here the bar means the time
averaging.  In contrast with Ref.\cite{KY} the field $A_{t}$
represents a noise with short-range time-correlations rather than
a strictly monochromatic field \cite{noise}.  The correlation
function $\overline{A_{t}\,A_{t'}}$ is supposed to decrease at
$|t-t'|>t_{c}\sim \omega^{-1}$.  For simplicity we consider this
field to be constant along a ring though the actual requirement
\cite{KY} for the scale $r_{c}$ of space variation is much weaker
$r_{c}\gg L_{\omega}$.

We consider two different geometries: a long wire with the
length ${\cal L}\gg L_{\varphi}$ and a ring with the
circumference $L\ll L_{\varphi}$.
In the latter case we study a DC current that flows when a {\it 
time-independent}
flux $\phi$ threads the ring.

The weak localization correction $\langle I_{wl}(t)\rangle$ to the
disorder-averaged current
in such a system is given by the well known cooperon contribution
\cite{AAK}:
\begin{equation}
\label{ccon}
\langle I_{wl}(t)\rangle= \frac{C_{\beta}\,e^2 D}{2\hbar
L}\int_{0}^{\infty}d\tau\,{\cal
C}_{t-\frac{\tau}{2}}
\left(\frac{\tau}{2},-\frac{\tau}{2}\right)\,E(t-\tau).
\end{equation}
Here ${\cal C}_{t}(\tau,\tau')=\sum_{q}{\cal C}_{t}(q,\tau,\tau')$ is a
cooperon at coincident space points, and ${\cal C}_{t}(q,\tau,\tau')$
is determined by \cite{AAK}:
\begin{equation}
\label{ceq}
\frac{\partial {\cal C}_{t}}{\partial \tau}+D\,\left(
q-\frac{e}{\hbar c}(A_{t+\tau}+A_{t-\tau})\right)^{2}{\cal
C}_{t}=\delta(\tau-\tau'),
\end{equation}
where $q$ is a momentum.  It is continuous if the wire is long;
for a ring $q=(2\pi/L)\,(m-2e\phi/hc),\;\;m=0,\pm 1,\pm 2...$.
Eqs.(\ref{ccon},\ref{ceq}) are valid if the conductance of the
system is large $g\gg 1$, and the field is weak enough $(e/\hbar
c)A_{t}l\ll 1$, ($l$ is the mean free path of the electrons).

The DC component of the current $ \langle I_{E}(\phi)\rangle $ is given by
the time
average of Eq.(\ref{ccon}).
Since this average does not depend on the
reference point we can shift $t\rightarrow t+\tau/2$ and express
the $n-th$ flux harmonic $I_{E}^{(n)}$ of the current
\begin{equation}
\label{suum}
\langle I_{E}(\phi)\rangle =\sum_{n=1}^{\infty}
I_{E}^{(n)}\,\sin(4\pi n\frac{ e\phi }{ hc }),
\end{equation}
through the $n-th$ flux harmonic
$C_{t}^{(n)}(\tau)$ of the cooperon 
\begin{equation}
\label{nth}
I_{E}^{(n)}= - \frac{i C_{\beta} e^2 D}{\hbar c
L}\int_{0}^{\infty}d\tau\,\overline{C_{t}^{(n)}(\tau)\,\frac{\partial
A_{t-\tau/2}}{\partial t}}.
\end{equation}
Solving Eq.(\ref{ceq}) and using the Poisson summation formula, one can
find an exact expression for $C_{t}^{(n)}(\tau)$:
\begin{equation}
\label{cnth}
C_{t}^{(n)}(\tau)=\sqrt{\frac{\tau_{D}}{4\pi\tau}}\,e^{-\frac{n^2
\tau_{D}}{4\tau}}\,e^{in\,S_{1}[A]}\,e^{-\tau\,S_{2}[A]}.
\end{equation}
Here
\begin{equation}
\label{s1}
S_{1}[A]=\frac{2eL}{\hbar
c}\,\left[\frac{1}{\tau}\int_{t-\tau/2}^{t+\tau/2}
A_{t_{1}}\,dt_{1}\right]\equiv \frac{2eL}{\hbar c}\, \langle
A_{t_{1}}\rangle_{t;\tau},
\end{equation}
\begin{equation}
\label{F}
S_{2}[A]=\frac{2e^2 D}{\hbar^{2}c^2}\,\left[\langle
A_{t_{1}}^{2}\rangle_{t;\tau}+\langle
A_{t_{1}}A_{2t-t_{1}}\rangle_{t;\tau}-2 \left(\langle
A_{t_{1}}\rangle_{t;\tau}\right)^{2} \right].
\end{equation}

According to Eqs.(\ref{ccon},\ref{ceq}) the weak-localization
correction to the conductance of a long wire equals to

\begin{equation}
\label{WLoc}
\delta\sigma= C_{\beta}\frac{\sqrt{\pi D}e^2}{2
h}\,\int_{0}^{\infty}\frac{d\tau}{\sqrt{\tau}}\,\overline{\exp\left\{
-\tau S_{2}[A]\right\}}.
\end{equation}
(we substitute a DC field $E_{0}$ for $E(t-\tau)$ and used the
definition Eq.(\ref{F})).  The form of
Eqs.(\ref{cnth},\ref{WLoc}) suggests that $S_{2}[A]$ is related
with the dephasing rate, while $S_{1}[A]$ is responsible for the
nonlinear DC current.

Now we assume that the correlation
time of the AC field is shorter than the relevant time scale
$\tau_{0}$ in the integrals Eqs.(\ref{nth},\ref{WLoc}).  For the
problem of dephasing in a long wire Eq.(\ref{WLoc})
$\tau_{0}^{(deph)}\sim \tau_{\varphi}$, while for the problem of
DC current in a ring Eq.(\ref{nth}) $\tau_{0}^{(DC)}\sim
\tau_{D}$.  Under these assumptions one can neglect the second
and the third terms in Eq.(\ref{F}) and identify the average
$\langle A_{t_{1}}^{2}\rangle_{t;\tau}$ defined in Eq.(\ref{s1})
with the true time-average $\overline{A_{t_{1}}^{2}}$. As a
result, $S_{2}[A]=2 D (e^2 /\hbar^{2}c^2)\,\overline{A^{2}_{t}}$
becomes independent of $t$ and $\tau$.  Using Eq.(\ref{WLoc}), we identify
$S_{2}$[A] with the noise-induced dephasing rate:
\begin{equation}
\label{dphr}
\frac{1}{\tau_{\varphi}}= 2 D (e^2 /\hbar^{2}c^2)\,\overline{A^{2}_{t}}.
\end{equation}

In order to compute the amplitude $I_{E}^{(n)}$ of the DC current we have
to evaluate the time-average in Eq.(\ref{nth})
\begin{eqnarray}
\label{aver}
\overline{\frac{\partial A_{t-\tau/2}}{\partial t}\,\exp\left\{ 
\frac{in}{\tau}\,
(2eL/\hbar c)\int_{t-\tau/2}^{t+\tau/2}\,
A_{t_{1}}\,dt_{1}
\right\}}=\\ \nonumber = \frac{i n}{\tau}\,
(2eL/\hbar c)\, \overline{A_{t}^{2}}.
\end{eqnarray}
Since the time-average of the total time-derivative is zero, we
can transfer the differentiation to the exponent. In the limit
$(L/L_{\varphi})^{2}\,(t_{c}/\tau_{D})\ll 1$, one can
differentiate only the lower limit of the integral and set
$\langle A_{t_{1}}\rangle_{t,\tau}=\overline{A_{t}}=0$ in the
exponent {\it after} the differentiation.  Substituting the
result in Eq.(\ref{nth}) we arrive at an integral over $\tau$,
which can be evaluated exactly. Finally, we use Eq.(\ref{dphr}) to express
$\overline{A_{t}^{2}}$ in terms of the dephasing rate and obtain
the amplitude of the $n-th$ flux harmonic of the DC
current averaged over the ensemble of mesoscopic rings
Eq.(\ref{suum}):
\begin{equation}
\label{fr}
I_{E}^{(n)}=C_{\beta}\,\left(\frac{e}{\tau_{\varphi}}
\right)\,\exp\left[-n\frac{L}{L_{\varphi}}\right].
\end{equation}
Eq.(\ref{WF}) for the principal $h/2e$- periodic component
$I_{E}^{(1)}$ is just in the limit of Eq.(\ref{fr}) at $L\ll L_{\varphi}$.

Unlike other theories \cite{AGef,AE} of persistent current, the
relationship Eq.(\ref{WF}) gives a correct magnitude of the DC
current.  Indeed, in a given sample at $T=0$ the current as a
function of the noise intensity reaches its maximal value when
$L_{\varphi}$ becomes comparable to the sample size $L$ (further
increase of the intensity would suppress the DC current
exponentially $\sim \exp\{-L/ L_{\varphi}\}$). This condition can be
rewritten as $\tau_{\varphi} \sim \tau_D$. Using Eq.(\ref{WF}) we
find \cite{KY} that the maximal value of the current is of the order of
\begin{equation}
\label{max}
I_{E}^{max} \sim e/\tau_D
\end{equation}
This is the order of magnitude of the current which was observed in all
experiments.

The ensemble-averaged current observed in copper rings by Levy et
al. \cite{Levy} was about 0.3 nA.
We can estimate $2e/\pi\tau_{\varphi} < $ 0.9 nA.
In Ref.\cite{Bouch} an ensemble of $10^5$ GaAs/GaAlAs rings
have been studied. In this case the estimation gives $4e/\pi
\tau_{\varphi}< 1.2$ nA, while the observed ensemble-averaged current was
about 1.5 nA.
In both cases there is a great deal of uncertainty: the saturated value of
$\tau_{\varphi}$ has not been measured, and for estimation we used values
of $\tau_{\varphi}$ measured in similar structures at $T\approx 1$ K and
$T\approx 50$ mK, respectively. Nevertheless the estimations of the DC
current based on Eq.(\ref{WF}) are much closer to the experimental values
than the predictions of the theories Ref.\cite{AGef}, which assume thermal
equilibrium.

Recently Mohanty et.al. measured the low-temperature dephasing
and the "persistent" current in the same set up \cite{MohW,Mo}.
The dephasing time in long gold wires saturated at
$\tau_{\varphi}\approx 4$ ns.  The "persistent" current has been
obtained from the magnetization of 30 gold rings fabricated in
the same way as the wires. The amplitude of the $h/2e$ DC
current was found to be $\sim$ 0.06 nA, while
$2e/\pi\tau_{\varphi}\approx$ 0.03 nA.  Therefore, in all three
experiments Refs.\cite{Levy,Bouch,MohW,Mo} Eq.(\ref{WF}) was
satisfied up to a factor $\sim 2$, though the magnitude of the
persistent current varied within two decades.

Situation with the sign of the magnetization is not so clear.  In
Ref.\cite{Levy} this sign was not measured.  In GaAs/GaAlAs rings
of Ref.\cite{Bouch}  the sign of $I_{E}$ was observed to
be {\it diamagnetic} in full agreement with Eq.(\ref{WF}) (the
spin-orbit effects are negligible  \cite{HB}).  However, in
Ref.\cite{MohW,Mo} the sign of the $h/2e$ response was also {\it
  diamagnetic}.  This contradicts Eq.(\ref{WF}): $I_{E}$ should
be {\it paramagnetic}, since spin-orbit effects in gold are
strong.  We believe, that the contradiction can be explained in
the following way.  The number of rings $N=30$ in
Ref.\cite{MohW,Mo} might be not sufficiently large to average out
the random mesoscopic part in the $h/2e$-periodic component of
the current.  Indeed, there can be no $h/e$-periodic component in
the {\it ensemble-averaged} current.  However, in
Ref.\cite{MohW,Mo} the $h/e$ magnetic response of 30 rings was only
twice as weak as the $h/2e$ one. This means that the
ensemble-averaged DC current $I_{E}$ and the $h/2e$-periodic
current averaged over 30 rings can differ substantially and even
have opposite signs.  The prediction that the sign of the
"persistent" current is determined by the strength of the
spin-orbit interaction is the most critical for our theory and
deserves a serious experimental check.

In conclusion, we have derived a relationship, Eq.(\ref{WF})
between the averaged DC current generated by a non-equilibrium AC
noise in an ensemble of mesoscopic rings and the dephasing rate
caused by the same noise. It provides a much better fit for the
magnitude of low-temperature ring magnetization than other
existing theories, which assume the equilibrium.  More
experimental work is needed to confirm the role of the
non-equilibrium noise.  However, there are reasons to suspect
that currently we deal with substantially non-equilibrium
mesoscopic systems.

We would like to thank I.L.Aleiner, C.Beenakker, H.Bouchiat, M.
Dyakonov, M.V.Feigelman, Yu.Galperin, K.B.Efetov, A.I.Larkin,
I.V.Lerner, P.Mohanty, V.I.Yudson and R.A.Webb for stimulating
discussions. The work at Princeton University was supported by ARO MURI
DAAG55-98-1-0270.

\widetext

\begin{references}
\bibitem{noise} This difference does not change the results of
  Ref.\cite{KY} for the DC current at $L\ll L_{\varphi}$ but it
  is relevant for the the dephasing rate in quasi-1D wires as
  well as for the DC current attenuation factor at
  $L>L_{\varphi}$.
\end{references}
\end{document}